\documentclass[11pt]{article}

\usepackage[margin=1in]{geometry}
\usepackage[T1]{fontenc}
\usepackage{cite}
\usepackage{amsmath,amssymb,amsfonts}
\usepackage{algorithmic}
\usepackage{graphicx}
\usepackage{textcomp}
\usepackage{subcaption}
\usepackage{siunitx}
\usepackage{multirow}
\usepackage[hidelinks]{hyperref}

\title{Efficient Spintronic Stochastic In-Sensor Computing Modules}
\author{%
Abdolah Amirany$^{1}$, HamidReza Imani$^{2}$, Samuel Farid$^{2}$,\\
Tarek El-Ghazawi$^{2}$\\[0.8em]
\small $^{1}$Herbert Wertheim College of Engineering, University of Florida,\\
\small Gainesville, FL 32611, USA\\[0.3em]
\small $^{2}$Department of Electrical and Computer Engineering,\\
\small The George Washington University, Washington, DC 20052, USA\\[0.5em]
\small\texttt{a\_amirany@ieee.org} \quad \texttt{hamidreza@gwu.edu}\\[0.2em]
\small\texttt{sfarid@gwu.edu} \quad \texttt{tarek@gwu.edu}
}
\date{}

\begin{document}
\maketitle

\begin{abstract}
In-sensor processing (ISP) is used in various fields, including autonomous driving automotive systems, wearable healthcare devices, environmental monitoring, wireless sensor networks, smart cities, and industrial control systems. One of the main features of most of these applications is energy and hardware restriction. This restriction amplifies the importance of using high-performance and low-energy consumption computing systems and methods. This paper proposes efficient spintronic stochastic computing modules for in-sensor computing systems to address this challenge. The proposed design utilized the stochastic behavior of the magnetic tunnel junction in the subcritical switching regime to generate a probability-adjustable random stream. Probability-adjustable random stream generation is the most important step in any stochastic system because it determines the quality of the final output. The simulation results indicate that the proposed design offers almost the same peak signal-to-noise ratio and structural similarity while occupying at least 62\% lower area and consuming 60\% lower energy. These advantages made the proposed design a good choice for ISP applications with limited energy.
\end{abstract}

\noindent\textbf{Keywords:} Edge Detection, In-Sensor Processing, Magnetic Tunnel Junction, Spintronic, Stochastic Computing.

\section{Introduction}
\label{sec:introduction}

In-sensor processing (ISP) offers a significant performance increase in very large scale integrated circuit (VLSI) design by combining data acquisition with computation ~\cite{zhou2020nearsensor,imani2024moe,imani2024rlfl}. This feature is crucial for modern applications that demand rapid, efficient, and intelligent decision-making ~\cite{suvizi2024exploring,lee2022insensor,wan2023insensor,kochakkashani2023supply,imani2022isample}. This fusion of sensing and processing allows for the reduction of data transmission loads, as it enables preprocessing of data right at the point of capture. ISP also enhances privacy and security by minimizing the amount of raw data that needs to be sent to centralized systems ~\cite{zhou2020nearsensor,mohseni2023adaptive,mohseni2024spectral}. By providing real-time insights, in-sensor processing is instrumental for applications in autonomous vehicles, smart cities, and internet of things  (IoT) devices, where the speed of response can be critical ~\cite{suvizi2025auto,zhou2020nearsensor,tang2020cim,alkabani2020oecam}. 

In energy-restricted systems, ISP is particularly transformative due to its ability to significantly reduce power consumption ~\cite{adelkhani2025lpcr,tang2020cim,mughrabi2024ecg}. By processing data locally within the sensor, there is a substantial decrease in the need to transmit large volumes of raw data to a central processor, which is a high-energy activity ~\cite{imani2022isample,anderson2023photonic}. This localized processing enables energy-restricted systems, such as battery-powered devices and remote sensors, to operate for extended periods without requiring frequent recharging or battery replacements ~\cite{zhou2020nearsensor,tang2020cim,baradaran2023cache}. Moreover, ISP can intelligently determine which data are essential and only send this filtered information for further analysis, leading to more energy-efficient data handling. This is vital for the proliferation of the IoT, where countless devices need to operate autonomously and continuously, often in settings where energy resources are limited, or the cost of energy consumption is a critical constraint ~\cite{suvizi2025parallel,bakhtiary2023sram,nazhamaiti2022vision}.

Spintronics, which exploits the spin property of electrons in addition to their charge, is grabbing significant attention in the field of ISP, particularly with the application of magnetic tunnel junction (MTJ) in random number generation ~\cite{rezaei2023spintronic,amirany2022spintronic,bahmanabadi2023flipflop}. Random number generation utilizing subcritical currents in MTJs involves the switching of magnetic layers by currents that are below the threshold (namely critical current) typically required to flip the magnetization ~\cite{khodayari2023trng1,khodayari2023trng2,amirany2021neuron}. This leads to stochastic switching behavior, which is unpredictable and ideal for generating random numbers. Since this process relies on thermal fluctuations assisted by smaller currents, it consumes less power, aligning well with the energy constraints of ISP systems ~\cite{amirany2021neuron}. In the context of ISP, spintronics also enhances the efficiency and speed of sensors by enabling nonvolatile memory that retains information without power ~\cite{zhou2020nearsensor,tang2020cim}. 

In this paper, by integration of spintronics into ISP, especially utilizing the subcritical current switching of MTJs, a probability adjustable stochastic number generator (SNG) is proposed. The proposed spintronics SNG offers great promises for efficient implementation of the stochastic systems as it offers low power consumption and area efficiency. 

The rest of the paper is organized as follows, the background of the paper including the spintronics technologies, MTJ and its stochastic behavior explained in Section~\ref{sec:background}. The proposed SNG, its application in ISP and especially in edge detection explained in Section~\ref{sec:proposed}. Simulation results of both the proposed SNG and stochastic edge detector are presented in section~\ref{sec:results}. Finally, based on the simulation results paper is concluded in Section~\ref{sec:conclusion}.

\section{Background}
\label{sec:background}
\subsection{MTJ and Its Stochastic Behavior}

\begin{figure}[t]
\centering

\begin{subfigure}{0.45\linewidth}
    \centering
    \includegraphics[width=\linewidth,height=4cm,keepaspectratio]{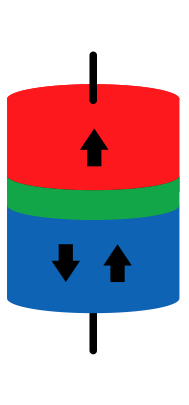}
    \caption{}
\end{subfigure}
\hfill
\begin{subfigure}{0.45\linewidth}
    \centering
    \includegraphics[width=\linewidth,height=4cm,keepaspectratio]{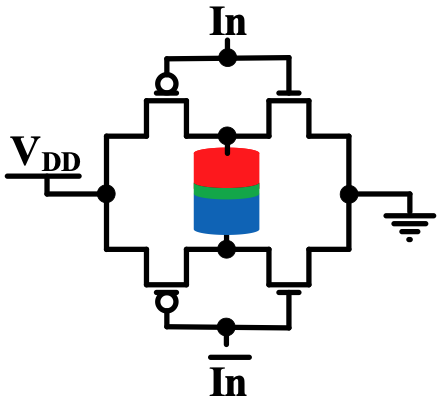}
    \caption{}
\end{subfigure}

\caption{MTJ Structure ~\cite{wang2014mtj} a) conceptual structure b) MTJ write circuit}
\label{fig:mtj}
\end{figure}

The MTJ is an essential element in spintronic circuits. It is a sandwich like structure of two ferromagnetic layers separated by a thin insulating barrier, as depicted in Fig.~\ref{fig:mtj}~\cite{wang2014mtj}. The functioning of an MTJ is set on a quantum phenomenon known as tunneling magnetoresistance. This phenomenon says that the resistance of the MTJ changes in accordance with the relative orientation of magnetic vector of the ferromagnetic layers~\cite{wang2014mtj}. In the case where these vectors are parallel ($P$), electron tunneling is facilitated, resulting in a low resistance ($R_P$). In contrast, when these vectors are antiparallel ($AP$), tunneling is less likely which results in a high resistance ($R_{AP}$)~\cite{wang2014mtj}.

A critical parameter within the MTJ device is the critical current ($I_{cri}$). $I_{cri}$ is defined as the minimal electrical current density required to grant a deterministic switching of the ferromagnetic layers within the specific time ($T_0$) ~\cite{wang2014mtj}. Below the critical current, MTJ switching will be probabilistic and follows the probability distribution expressed by Eq.~\ref{eqn:prob_dest}.

\begin{equation}
\label{eqn:prob_dest}
P(I,t) = 1 - \exp\left\{ -\frac{t}{T_0} \exp\left[ -\Delta \left(1 - \frac{I}{I_{\mathrm{cri}}} \right)^2 \right] \right\}
\end{equation}

\begin{figure}[t]
\includegraphics[width=\linewidth]{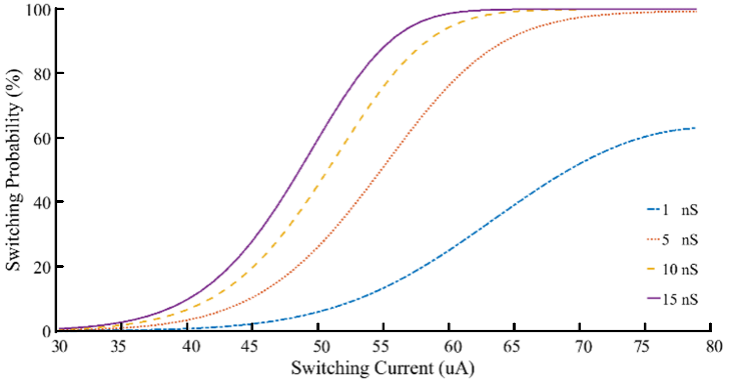}
\caption{MTJ switching probability for different current pulse duration}
\label{fig:mtj_prob}

\end{figure}
In Eq.~\ref{eqn:prob_dest}, $t$ is the width of the applied current pulse, $\delta$ is the thermal stability factor, and  $I$ is the current magnitude pass through MTJ ~\cite{qu2017trng}. Figure ~\ref{fig:mtj_prob} shows the MTJ switching probability for different current pulse duration~\cite{wang2014mtj}. This feature has many applications including random number generation~\cite{mehri2022rng}, stochastic neural network~\cite{amirany2021neuron}, and approximate memories~\cite{ranjan2015approx}. One of the fields where this feature can be very helpful is stochastic computing (SC)~\cite{zink2022review,zink2023scram}.

\begin{figure}
    \centering
    \includegraphics[width=\linewidth]{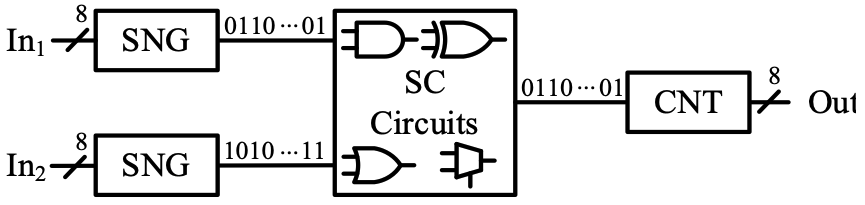}
    \caption{Overview of an SC systems~\cite{gross2019stochastic}}
    \label{fig:overview_SC}
\end{figure}

\subsection{Stochastic Computing}
SC emerges as a paradigm that fundamentally deviates from conventional binary computing~\cite{gross2019stochastic,salehi2020rng}. In SC, numbers are representing and manipulating data in the form of stochastic bitstreams~\cite{gross2019stochastic}. This form of computing is inherently robust to soft errors and exhibits lower power consumption. The main application of SC is in the domain of Digital image and signal processing and artificial intelligence~\cite{gross2019stochastic,salehi2020rng,alaghi2013stochastic,alaghi2012spectral}. 
An SC system as shown in Fig.~\ref{fig:overview_SC} is comprised of three steps~\cite{gross2019stochastic}. At its core, stochastic computing relies on the generation of random numbers in SNG, which serves as the backbone for constructing stochastic representations of data. The quality of these random numbers is critical, as it directly impacts the accuracy and reliability of the stochastic computations~\cite{gross2019stochastic,salehi2020rng}.

The second stage involves the manipulation of these stochastic bitstreams within stochastic circuits. These circuits perform arithmetic operations by simple logic gates that operate on bitstreams, exploiting the randomness to encode and process information. For instance, multiplication can be performed by a simple AND gate, and addition by a multiplexer, which contrasts completely with the complex circuitry required in binary systems \cite{gross2019stochastic}. The simplicity of such operations translates into smaller and more energy-efficient circuit designs. Technologies employing devices like MTJ are at the forefront of the second stage, leveraging the subcritical current switching phenomenon to produce high-quality random bitstreams while maintaining low power profiles. These bitstreams are the probabilistic analogs of binary values, with the ratio of ones to zeros representing a real number \cite{gross2019stochastic}.

Finally, to retrieve usable outputs from the stochastic domain, a counter (CNT) or a similar circuit is employed at the last stage to convert the stochastic bitstream back into a binary number \cite{gross2019stochastic}.

\section{Proposed SNG and its applications}
\label{sec:proposed}
In this section, first we explain the structure and operation of the proposed spintronic SNG and then present the proposed stochastic edge detection circuits for ISP applications.  

\begin{figure}
    \centering
    \includegraphics[width=1\linewidth]{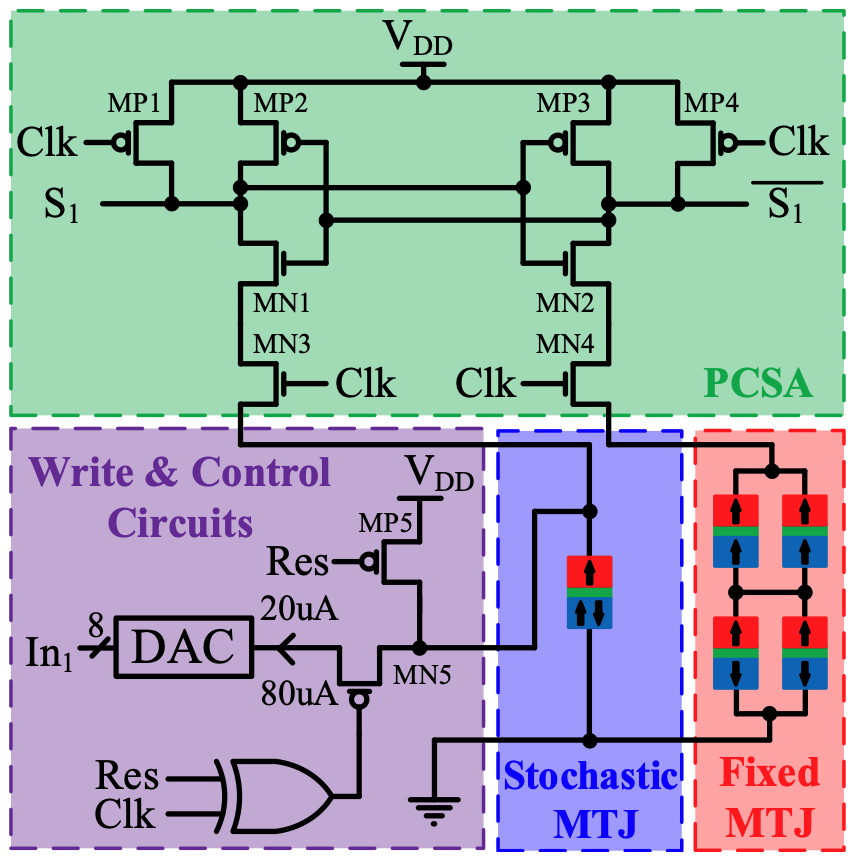}
    \caption{Proposed Spintronic SNG}
    \label{fig:proposed_SNG}
\end{figure}

\subsection{Proposed Spintronic SNG}
Figure~\ref{fig:proposed_SNG} depicts the proposed spintronic SNG. As shown in this figure, the proposed spintronic SNG consists of four primary parts:
\subsubsection{Precharge Sense Amplifier}
The SA is responsible for detecting resistance differences between two MTJs connected to its tails. Among different SA types, the PCSA stands out due to its superior sensing reliability, performance, and power.

\begin{figure}
    \centering
    \includegraphics[width=1\linewidth]{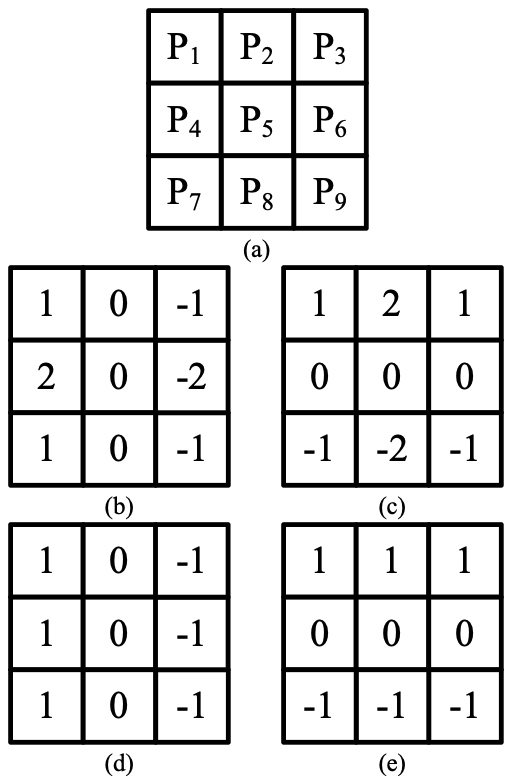}
    \caption{Sobel and Prewitt masks a) $3\times3$ input image b) vertical Sobel mask c) horizontal Sobel mask d) vertical Prewitt mask e) horizontal Prewitt mask}
    \label{fig:masks}
\end{figure}

As depicted in Fig.~\ref{fig:proposed_SNG}, the PCSA consists of eight transistors, including two PMOS transistors ($MP1$ and $MP4$) serving as precharge paths, four NMOS/PMOS transistors ($MN1$, $MN2$, $MP2$, and $MP3$) forming two cross-coupled inverters to create a keeper and two NMOS transistors ($MN3$ and $MN4$) serving as discharge path. The PCSA's sensing operation encompasses two phases:
\begin{enumerate}
    \item Pre-charge phase (Clk=‘0’): During this phase, both output nodes are pre-charged to $V_{DD}$ through $MP1$ and $MP4$. Simultaneously, write and control circuits execute MTJ reconfiguration.
    \item Evaluation phase (Clk=‘1’): In this phase, $MN3$ and $MN4$ turn on to discharge one of the outputs to the ground. The MTJs with lower resistance will discharge its corresponding output to the ground faster, causing its complementary output to remain high.
\end{enumerate}

\subsubsection{Write and Control Circuits}
The write and control circuits are responsible for reconfiguring the stochastic MTJ. The operation of this part of the proposed spintronic SNG can be divided into two phases:
\begin{enumerate}
    \item Deterministic restoring ($Clk=0$, and $Res=0$): During this phase, $MP5$ turns on, allowing a current of 200$\mu$A to pass through the stochastic MTJ for 10ns. This current, which exceeds the MTJ's critical current (100$\mu$A), definitively changes the MTJ's state to antiparallel.
    \item Stochastic switching ($Clk=0$, and $Res=1$): In this phase, a current ranging from -20 to -80$\mu$A flows through the stochastic MTJ for 10ns. During this time, the state of the stochastic MTJ can change according to the probability distribution expressed by Eq.~\ref{eqn:prob_dest}.
\end{enumerate}

\begin{figure}
    \centering
    \includegraphics[width=1\linewidth]{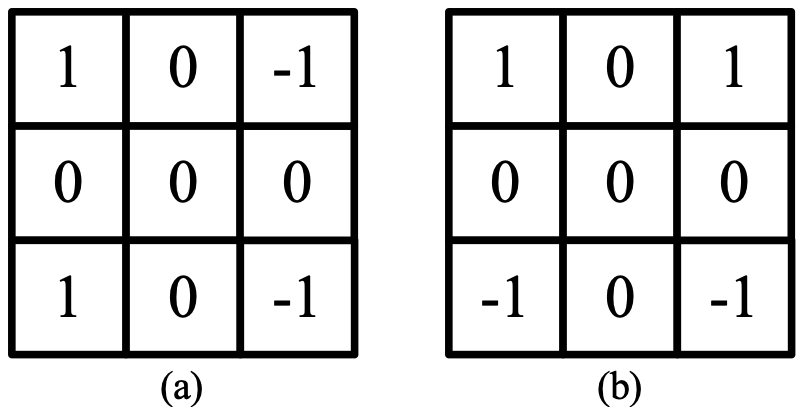}
    \caption{Proposed stochastic masks a) vertical mask b) horizontal mask}
    \label{fig:modfied_masks}
\end{figure}

\begin{figure}
    \centering
    \includegraphics[width=1\linewidth]{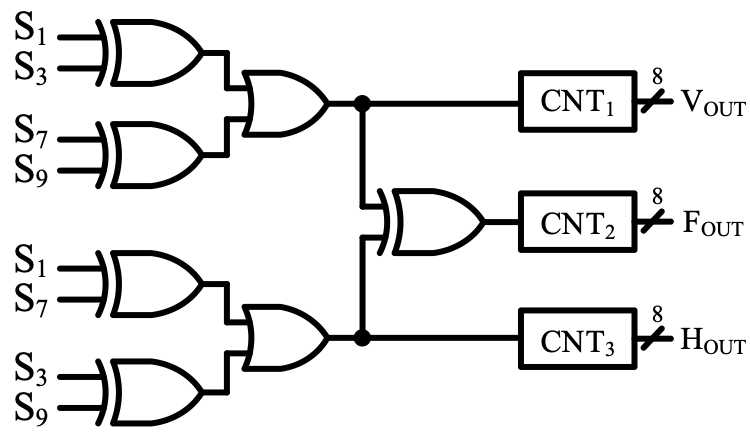}
    \caption{Implementation of the proposed masks}
    \label{fig:implementation}
\end{figure}

\subsubsection{Stochastic MTJ}
This circuit part contains an stochastic MTJ and is responsible for introducing the stochastic nature to the proposed spintronic SNG.
\subsubsection{Fixed MTJs}
This part of the proposed spintronic SNG integrates four fixed MTJs, including two parallel MTJs and two antiparallel MTJs, with an equivalent resistance of ($(R_P+R_{AP})/2$). This part serves as a reference resistor for comparing the resistance of the stochastic MTJ.
\subsection{Proposed Stochastic Edge Detector}
Edge detection in ISP is critical for reducing data transmission and power consumption in applications like image processing and autonomous systems. By identifying significant changes in pixel values, ISP isolates key features directly within the sensor, enhancing processing speed and efficiency. This is particularly beneficial in power-limited environments, aiding real-time applications such as surveillance and autonomous navigation, where swift and efficient data processing is essential.

Sobel and Prewitt are two of the well-known and widely used edge detection algorithms. Figure~\ref{fig:masks} shows the vertical and horizontal mask of the Sobel and Prewitt filters. To simultaneously extract the horizontal and vertical edges, it is enough to apply two masks consecutively on the image. Also, due to the linear nature of the convolution operation, it is possible to multiply the mask of vertical and horizontal filters and apply the resulting mask to the image.

Based on the Prewitt mask, two modified masks are proposed in this paper. These masks are shown in Fig.~\ref{fig:modfied_masks}. Compared to the original Prewitt masks (Fig.~\ref{fig:masks}.d and~\ref{fig:masks}.e), in these masks two coefficients are removed. The reason for removing these two coefficients is to prevent detection of false edges, because false edges are usually two consecutive pixels. Another noteworthy point is that although these two masks are designed for stochastic implementation, they can also be used in conventional implementation, and considering that only coefficients ‘1’ and ‘-1’ are used, conventional implementation will also be very simple and efficient.

Figure ~\ref{fig:implementation} shows the implementation of the proposed mask in stochastic domain. As this figure shows, using proposed implementation, vertical, horizontal, and full edge can be detected simultaneously. 

\section{Simulation Results}
\label{sec:results}

\begin{table}[t]
\caption{Parameters of the CMOS Transistors and the MTJ Devices and Their Variations}
\label{tab:params}
\centering
\resizebox{\columnwidth}{!}{
\begin{tabular}{|l|l|c|c|}
\hline
\multicolumn{2}{|c|}{\textbf{Parameter}} & \textbf{Value} & \textbf{$\pm$3$\sigma$ Variation} \\
\hline

\multicolumn{4}{|c|}{\textbf{MTJ}} \\
\hline

\multicolumn{2}{|l|}{Shape} & \multicolumn{2}{c|}{Round} \\
\hline

\multicolumn{2}{|l|}{MTJ diameter} & 40 nm & 7.2\% \\
\multicolumn{2}{|l|}{Barrier thickness} & 0.9 nm & 5\% \\
\multicolumn{2}{|l|}{Resistance area product} & $10\,\Omega\cdot\mu m^2$ & 15\% \\
\multicolumn{2}{|l|}{Free layer thickness} & 0.9 nm & 5\% \\
\multicolumn{2}{|l|}{TMR under zero bias voltage ($\mathrm{TMR}_0$)} & 200\% & 10\% \\
\multicolumn{2}{|l|}{MTJ resistance} & $10 \sim 30$ k$\Omega$ & -- \\
\multicolumn{2}{|l|}{Switching time} & $\sim$5 ns & -- \\
\hline

\multicolumn{4}{|c|}{\textbf{Transistors}} \\
\hline

\multicolumn{2}{|l|}{Gate length} & 40 nm & 5\% \\
\hline

\multirow{2}{*}{Gate width} & Minimum size & 40 nm & 10\% \\
 & Conventional write circuit & 400 nm & -- \\
\hline

\multicolumn{4}{|c|}{\textbf{Simulation Setup}} \\
\hline

\multicolumn{2}{|l|}{Supply voltage} & 1.0 V & 10\% \\
\multicolumn{2}{|l|}{Temperature} & 300 K & 270 K -- 360 K \\
\multicolumn{2}{|l|}{Simulation step} & 10 ps & -- \\
\multicolumn{2}{|l|}{Clock frequency} & 100 MHz & -- \\
\hline

\end{tabular}
}
\end{table}

\begin{table}[t]
\caption{Performance of the Proposed Approximate Spintronic Memory}
\label{tab:spintronic_perf}
\centering
\resizebox{\textwidth}{!}{%
\begin{tabular}{|l|l|l|c|c|c|c|c|}
\hline
\multicolumn{3}{|c|}{\textbf{Design}} & \textbf{Energy consumption} & \multicolumn{2}{c|}{\textbf{Area}} & \textbf{E-PSNR} & \textbf{E-SSIM} \\
\cline{5-6}
\multicolumn{3}{|c|}{} &  & \textbf{E-NAND} & \textbf{MTJ} &  &  \\
\hline

\multicolumn{3}{|l|}{Canny Edge Detector (Ground Truth)} 
& 3675.2 & $\sim$30,000 & 0 & $\infty$ & 1.0000 \\
\hline

\multirow{3}{*}{\shortstack{Sobel\\Edge Detector}}
& \multicolumn{2}{l|}{Horizontal} & 124.3 & 964  & 0 & 46.88 & 0.9701 \\
\cline{2-8}
& \multicolumn{2}{l|}{Vertical}   & 121.3 & 964  & 0 & 50.01 & 0.9809 \\
\cline{2-8}
& \multicolumn{2}{l|}{Full}       & 181.0 & 1492 & 0 & 56.83 & 0.9953 \\
\hline

\multirow{3}{*}{\shortstack{Prewitt\\Edge Detector}}
& \multicolumn{2}{l|}{Horizontal} & 85.5  & 700 & 0 & 50.05 & 0.9810 \\
\cline{2-8}
& \multicolumn{2}{l|}{Vertical}   & 87.9  & 700 & 0 & 51.12 & 0.9809 \\
\cline{2-8}
& \multicolumn{2}{l|}{Full}       & 123.7 & 964 & 0 & 58.09 & 0.9961 \\
\hline

\multicolumn{3}{|l|}{Proposed in~\cite{alaghi2013stochastic}} 
& 39.1 & 336 & 0 & 57.21 & 0.9960 \\
\hline

\multicolumn{3}{|l|}{Proposed in~\cite{li2011image}} 
& 78.1 & 656 & 0 & 56.12 & 0.9958 \\
\hline

\multirow{9}{*}{\parbox{2.6cm}{\centering Proposed\\Stochastic\\Edge Detector\\in This Paper}}
& \multirow{3}{*}{Horizontal}
& 64 Bits  & 29.3 & 226 & \multirow{9}{*}{2} & 47.21 & 0.9721 \\
\cline{3-5}\cline{7-8}
& & 128 Bits & 30.7 & 239 &  & 50.12 & 0.9801 \\
\cline{3-5}\cline{7-8}
& & 256 Bits & 31.1 & 252 &  & 52.11 & 0.9844 \\
\cline{2-5}\cline{7-8}

& \multirow{3}{*}{Vertical}
& 64 Bits  & 28.7 & 226 &  & 49.81 & 0.9765 \\
\cline{3-5}\cline{7-8}
& & 128 Bits & 31.7 & 239 &  & 51.31 & 0.9812 \\
\cline{3-5}\cline{7-8}
& & 256 Bits & 32.9 & 252 &  & 54.12 & 0.9862 \\
\cline{2-5}\cline{7-8}

& \multirow{3}{*}{Full}
& 64 Bits  & 30.8 & 233 &  & 57.45 & 0.9961 \\
\cline{3-5}\cline{7-8}
& & 128 Bits & 29.9 & 247 &  & 58.11 & 0.9963 \\
\cline{3-5}\cline{7-8}
& & 256 Bits & 34.9 & 268 &  & 58.46 & 0.9970 \\
\hline

\end{tabular}%
}
\end{table}
In this section first, we present the simulation results of the proposed spintronic SNG and investigate its application in the edge detection. Simulations are performed in HSPICE simulator and using MTJ model proposed in~\cite{wang2014mtj}. Table~\ref{tab:params} shows the critical parameter of the MTJs and CMOS transistors. It should be noted that in all the circuit-level simulations, the effects of the fabrication process variation are also considered. For this purpose, the sensitive parameters of the circuit have been selected using a normal distribution and variance mentioned in Table~\ref{tab:params}.

\subsection{Functional Simulation}

\begin{figure}
    \centering
    \includegraphics[width=1\linewidth]{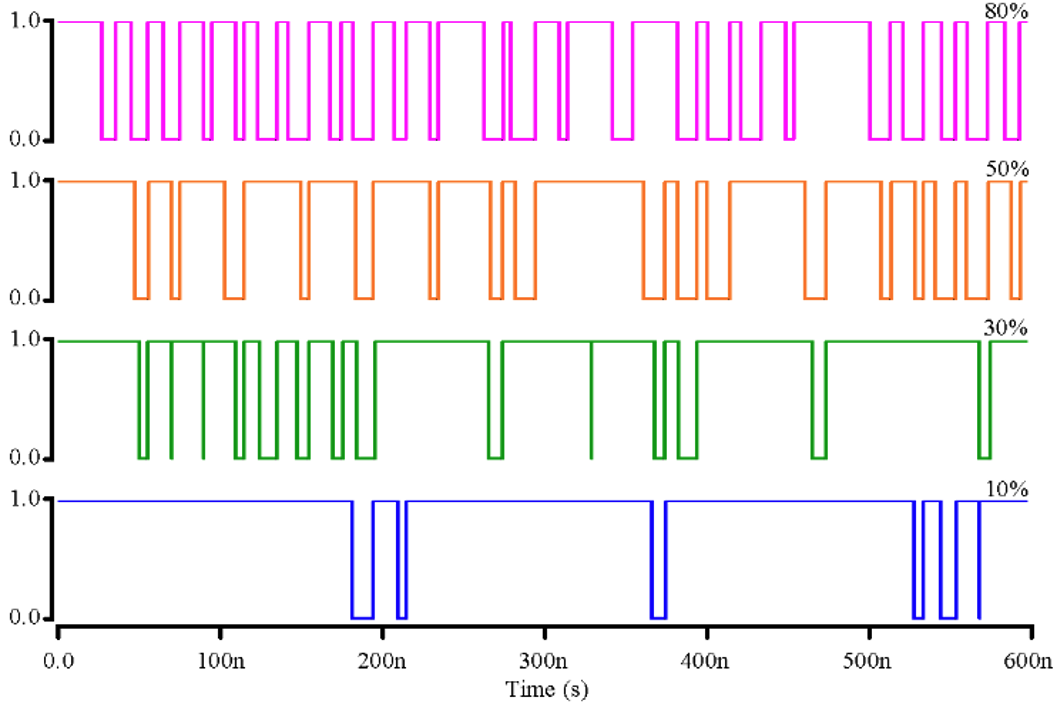}
    \caption{Transient simulation results of the proposed SNG}
    \label{fig:sim_results}
\end{figure}

To check the effectiveness of the proposed spintronic SNG circuit in generating random numbers with an adjustable ratio, a thousand samples of the output were generated for different MTJ switching currents. Figure ~\ref{fig:sim_results} shows four examples of generated outputs with corresponding probabilities. As Fig. 8 shows, by applying the appropriate current controlled by the control circuit, a random string with a fully adjustable ratio can be generated.

\subsection{Edge Detection Application}

To investigate the application of the proposed spintronic SNG in the edge detection following steps were performed using MATLAB:
\begin{enumerate}
    \item Each pixel is converted to a random string with adjustable ratio and length based on the results of circuit-level simulations. At this stage, the effect of fabrication process variation is also considered on the quality of the generated stream so that the results are completely realistic.
    \item The generated streams are applied to the proposed edge detection circuit.
    \item The number of ‘1’ in the output of the edge detection circuit is counted and converted into an 8-bit number.
    \item The output is binarized by applying an appropriate threshold value.
    \item The output image is compared with the output of Canny edge detector \cite{canny1986edge} as ground truth with the help of edge pick signal to noise ratio (E-PSNR) and edge structural similarity (E-SSIM).
\end{enumerate}

\begin{figure}
    \centering
    \includegraphics[width=1\linewidth]{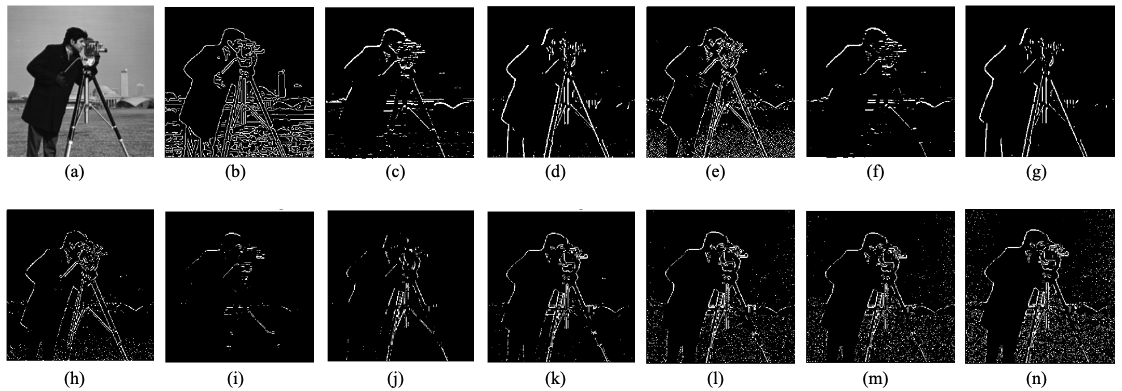}
    \caption{Results of the implemented edge detector a) original image b) output of Canny edge detector c) output of horizontal Sobel d) output of vertical Sobel    e) output of full Sobel f) output of horizontal Prewitt g) output of vertical Prewitt h) output of full Prewitt i) output of proposed horizontal edge detector with 256 bits stream j) output of proposed vertical edge detector with 256 bits stream k) output of proposed full edge detector with 256 bits stream l) output of proposed full edge detector with 128 bits stream m) output of proposed full edge detector with 64 bits stream n) output of proposed full edge detector with 32 bits stream}
    \label{fig:results_cameraman}
\end{figure}

Figure~\ref{fig:results_cameraman} shows the results of our proposed edge detection algorithms alongside those of the Sobel, Prewitt, and Canny edge detectors. As this figure shows, the proposed method has a visual output similar to and even better than the Soble and Prewitt methods. Another interesting thing in Fig.~\ref{fig:results_cameraman} is the acceptable output of the proposed method with 32-, 64-, and 128-bits stream. This feature is worthy of attention in ISP because the number of performed operations can be reduced by reducing the number of stream bits which results in lower power consumption. Also, this feature increases the overall system speed.

Table~\ref{tab:spintronic_perf} shows the results of the proposed stochastic edge detector compared to the Soble and Prewitt edge detection algorithms and stochastic edge detector proposed in~\cite{alaghi2013stochastic,li2011image}. As this table demonstrates, the proposed method outperforms the stochastic edge detectors proposed in~\cite{wang2014mtj,ranjan2015approx} in terms of energy consumption and area by 10$\%$ and 20$\%$ respectively. Compared to the Soble and Prewitt edge detection algorithms, the proposed method in this paper, even with 128-bits stream, offers same E-PSNR and E-SSIM while consuming 82$\%$ lower power and occupies 74$\%$ lower area.

\section{Conclusion}
\label{sec:conclusion}

This paper proposed a spintronic stochastic computing module for energy restricted applications such as autonomous vehicles, wearable healthcare devices, environmental monitoring, and smart city infrastructure. The proposed module utilizes the stochastic behavior of MTJ device in the subcritical switching regime, enabling the generation of probability-adjustable random streams which are crucial for the quality of stochastic systems. The simulation results in the edge detection application show that the proposed design maintains comparable E-PSNR and E-SSIM to existing methods, while significantly reducing area by at least 62$\%$ and energy consumption by 60$\%$. These attributes position the proposed design as a suitable choice for ISP applications where energy efficiency and low cost design are essential.

\bibliographystyle{IEEEtran}
\bibliography{refs}

@article{zhou2020nearsensor,
  title={Near-sensor and in-sensor computing},
  author={Zhou, Feichi and Chai, Yang},
  journal={Nature Electronics},
  volume={3},
  number={11},
  pages={664--671},
  year={2020},
  publisher={Nature Publishing Group UK London}
}

@INPROCEEDINGS{imani2024moe,
  author={Imani, HamidReza and Amirany, Abdolah and El-Ghazawi, Tarek},
  booktitle={2024 IEEE International Conference on Rebooting Computing (ICRC)}, 
  title={Mixture of Experts with Mixture of Precisions for Tuning Quality of Service}, 
  year={2024},
  volume={},
  number={},
  pages={1-6},
  doi={10.1109/ICRC64395.2024.10937027}}

@ARTICLE{imani2024rlfl,
  author={Imani, HamidReza and Anderson, Jeff and Farid, Samuel and Amirany, Abdolah and El-Ghazawi, Tarek},
  journal={IEEE Journal on Emerging and Selected Topics in Circuits and Systems}, 
  title={RLFL: A Reinforcement Learning Aggregation Approach for Hybrid Federated Learning Systems Using Full and Ternary Precision}, 
  year={2024},
  volume={14},
  number={4},
  pages={673-687},
  doi={10.1109/JETCAS.2024.3483554}}

@article{lee2022insensor,
  title={In-sensor image memorization and encoding via optical neurons for bio-stimulus domain reduction toward visual cognitive processing},
  author={Lee, Doeon and Park, Minseong and Baek, Yongmin and Bae, Byungjoon and Heo, Junseok and Lee, Kyusang},
  journal={Nature Communications},
  volume={13},
  number={1},
  pages={5223},
  year={2022},
  publisher={Nature Publishing Group UK London}
}

@article{wan2023insensor,
  title={In-sensor computing: materials, devices, and integration technologies},
  author={Wan, Tianqing and Shao, Bangjie and Ma, Sijie and Zhou, Yue and Li, Qiao and Chai, Yang},
  journal={Advanced materials},
  volume={35},
  number={37},
  pages={2203830},
  year={2023},
  publisher={Wiley Online Library}
}

@article{kochakkashani2023supply,
  title={Supply chain planning of vaccine and pharmaceutical clusters under uncertainty: The case of COVID-19},
  author={Kochakkashani, Farid and Kayvanfar, Vahid and Haji, Alireza},
  journal={Socio-economic planning sciences},
  volume={87},
  pages={101602},
  year={2023},
  publisher={Elsevier}
}

@inproceedings{imani2022isample,
  title={iSample: Intelligent client sampling in federated learning},
  author={Imani, HamidReza and Anderson, Jeff and El-Ghazawi, Tarek},
  booktitle={2022 IEEE 6th International Conference on Fog and Edge Computing (ICFEC)},
  pages={58--65},
  year={2022},
  organization={IEEE}
}

@inproceedings{mohseni2023adaptive,
  title={Adaptive conditional quantile neural processes},
  author={Mohseni, Peiman and Duffield, Nick and Mallick, Bani and Hasanzadeh, Arman},
  booktitle={Uncertainty in Artificial Intelligence},
  pages={1445--1455},
  year={2023},
  organization={PMLR}
}

@article{mohseni2024spectral,
  title={Spectral convolutional conditional neural processes},
  author={Mohseni, Peiman and Duffield, Nick},
  journal={arXiv preprint arXiv:2404.13182},
  year={2024}
}

@INPROCEEDINGS{tang2020cim,
  author={Tang, Kea-Tiong and Wei, Wei-Chen and Yeh, Zuo-Wei and Hsu, Tzu-Hsiang and Chiu, Yen-Cheng and Xue, Cheng-Xin and Kuo, Yu-Chun and Wen, Tai-Hsing and Ho, Mon-Shu and Lo, Chung-Chuan and Liu, Ren-Shuo and Hsieh, Chih-Cheng and Chang, Meng-Fan},
  booktitle={2019 Symposium on VLSI Circuits}, 
  title={Considerations Of Integrating Computing-In-Memory And Processing-In-Sensor Into Convolutional Neural Network Accelerators For Low-Power Edge Devices}, 
  year={2019},
  volume={},
  number={},
  pages={T166-T167},
  doi={10.23919/VLSIC.2019.8778074}}

@article{alkabani2020oecam,
  title={OE-CAM: a hybrid opto-electronic content addressable memory},
  author={Alkabani, Yousra and Miscuglio, Mario and Sorger, Volker J and El-Ghazawi, Tarek},
  journal={IEEE Photonics Journal},
  volume={12},
  number={2},
  pages={1--14},
  year={2020},
  publisher={IEEE}
}

@inproceedings{mughrabi2024ecg,
  title={ECG: Expressing locality and prefetching for optimal caching in graph structures},
  author={Mughrabi, Abdullah T and Baradaran, Morteza and Samara, Ahmed and Skadron, Kevin},
  booktitle={2024 IEEE International Parallel and Distributed Processing Symposium Workshops (IPDPSW)},
  pages={520--525},
  year={2024},
  organization={IEEE}
}

@article{anderson2023photonic,
  title={Virtualizing a post-moore’s law analog mesh processor: The case of a photonic pde accelerator},
  author={Anderson, Jeff and Kayraklioglu, Engin and Reza Imani, Hamid and Shen, Chen and Miscuglio, Mario and Sorger, Volker J and El-Ghazawi, Tarek},
  journal={ACM Transactions on Embedded Computing Systems},
  volume={22},
  number={2},
  pages={1--26},
  year={2023},
  publisher={ACM New York, NY}
}

@inproceedings{baradaran2023cache,
  title={Energy consumption analysis of instruction cache prefetching methods},
  author={Baradaran, Morteza and Ansari, Ali and Sadrosadati, Mohammad and Sarbazi-Azad, Hamid},
  booktitle={2023 International Symposium on Computer Architecture and High Performance Computing Workshops (SBAC-PADW)},
  pages={60--67},
  year={2023},
  organization={IEEE}
}

@article{bakhtiary2023sram,
  title={An SEU-hardened ternary SRAM design based on efficient ternary C-elements using CNTFET technology},
  author={Bakhtiary, Vahid and Amirany, Abdolah and Moaiyeri, Mohammad Hossein and Jafari, Kian},
  journal={Microelectronics Reliability},
  volume={140},
  pages={114881},
  year={2023},
  publisher={Elsevier}
}

@inproceedings{nazhamaiti2022vision,
  title={In-situ self-powered intelligent vision system with inference-adaptive energy scheduling for BNN-based always-on perception},
  author={Nazhamaiti, Maimaiti and Su, Haijin and Xu, Han and Liu, Zheyu and Qiao, Fei and Wei, Qi and Du, Zidong and Yang, Xinghua and Luo, Li},
  booktitle={Proceedings of the 59th ACM/IEEE Design Automation Conference},
  pages={913--918},
  year={2022}
}

@article{rezaei2023spintronic,
  title={A high-capacity and nonvolatile spintronic associative memory hardware accelerator},
  author={Rezaei, Mahan and Amirany, Abdolah and Moaiyeri, Mohammad Hossein and Jafari, Kian},
  journal={IET Circuits, Devices \& Systems},
  volume={17},
  number={4},
  pages={205--212},
  year={2023},
  publisher={Wiley Online Library}
}

@article{amirany2022spintronic,
  title={Highly reliable bio-inspired spintronic/CNTFET multi-bit per cell nonvolatile memory},
  author={Amirany, Abdolah and Jafari, Kian and Moaiyeri, Mohammad Hossein},
  journal={AEU-International Journal of Electronics and Communications},
  volume={158},
  pages={154452},
  year={2023},
  publisher={Elsevier}
}

@inproceedings{bahmanabadi2023flipflop,
  title={Towards nonvolatile spintronic quaternary flip-flop and register design},
  author={BahmanAbadi, Motahareh and Amirany, Abdolah and Moaiyeri, Mohammad Hossein and Jafari, Kian},
  booktitle={Spin},
  volume={13},
  number={03},
  pages={2350015},
  year={2023},
  organization={World Scientific}
}

@article{khodayari2023trng1,
  title={A variation-aware ternary true random number generator using magnetic tunnel junction at subcritical current regime},
  author={Khodayari, Fatemeh and Amirany, Abdolah and Moaiyeri, Mohammad Hossein and Jafari, Kian},
  journal={IEEE Transactions on Magnetics},
  volume={59},
  number={3},
  pages={1--8},
  year={2023},
  publisher={IEEE}
}

@article{khodayari2023trng2,
  title={Low-cost and variation-aware spintronic ternary random number generator},
  author={Khodayari, Fatemeh and Amirany, Abdolah and Jafari, Kian and Moaiyeri, Mohammad Hossein},
  journal={Circuits, Systems, and Signal Processing},
  volume={43},
  number={2},
  pages={1175--1191},
  year={2024},
  publisher={Springer}
}

@inproceedings{amirany2021neuron,
  title={Stochastic spintronic neuron with application to image binarization},
  author={Amirany, Abdolah and Meghdadi, Masoud and Moaiyeri, Mohammad Hossein and Jafari, Kian},
  booktitle={2021 26th International Computer Conference, Computer Society of Iran (CSICC)},
  pages={1--5},
  year={2021},
  organization={IEEE}
}

@article{wang2014mtj,
  title={Compact model of magnetic tunnel junction with stochastic spin transfer torque switching for reliability analyses},
  author={Wang, You and Zhang, Yue and Deng, EY and Klein, Jacques-Olivier and Naviner, Lirida AB and Zhao, WS},
  journal={Microelectronics Reliability},
  volume={54},
  number={9-10},
  pages={1774--1778},
  year={2014},
  publisher={Elsevier}
}

@inproceedings{qu2017trng,
  title={A true random number generator based on parallel STT-MTJs},
  author={Qu, Yuanzhuo and Han, Jie and Cockburn, Bruce F and Pedrycz, Witold and Zhang, Yue and Zhao, Weisheng},
  booktitle={Design, Automation \& Test in Europe Conference \& Exhibition (DATE), 2017},
  pages={606--609},
  year={2017},
  organization={IEEE}
}

@article{mehri2022rng,
  title={Theoretical circuit design of an efficient spintronic random number generator with an internal postprocessing unit},
  author={Mehri, Saeed and Amirany, Abdolah and Moaiyeri, Mohammad Hossein and Jafari, Kian},
  journal={IEEE Magnetics Letters},
  volume={13},
  pages={1--5},
  year={2022},
  publisher={IEEE}
}

@inproceedings{ranjan2015approx,
  title={Approximate storage for energy efficient spintronic memories},
  author={Ranjan, Ashish and Venkataramani, Swagath and Fong, Xuanyao and Roy, Kaushik and Raghunathan, Anand},
  booktitle={Proceedings of the 52nd Annual Design Automation Conference},
  pages={1--6},
  year={2015}
}

@article{zink2022review,
  title={Review of magnetic tunnel junctions for stochastic computing},
  author={Zink, Brandon R and Lv, Yang and Wang, Jian-Ping},
  journal={IEEE Journal on Exploratory Solid-State Computational Devices and Circuits},
  volume={8},
  number={2},
  pages={173--184},
  year={2022},
  publisher={IEEE}
}

@article{zink2023scram,
  title={A stochastic computing scheme of embedding random bit generation and processing in computational random access memory (SC-CRAM)},
  author={Zink, Brandon R and Lv, Yang and Zabihi, Masoud and Cilasun, Husrev and Sapatnekar, Sachin S and Karpuzcu, Ulya R and Riedel, Marc D and Wang, Jian-Ping},
  journal={IEEE Journal on Exploratory Solid-State Computational Devices and Circuits},
  volume={9},
  number={1},
  pages={29--37},
  year={2023},
  publisher={IEEE}
}

@book{gross2019stochastic,
  title={Stochastic computing: techniques and applications},
  author={Gross, Warren J and Gaudet, Vincent C},
  year={2019},
  publisher={Springer}
}

@article{salehi2020rng,
  title={Low-cost stochastic number generators for stochastic computing},
  author={Salehi, Sayed Ahmad},
  journal={IEEE Transactions on Very Large Scale Integration (VLSI) Systems},
  volume={28},
  number={4},
  pages={992--1001},
  year={2020},
  publisher={IEEE}
}

@inproceedings{alaghi2013stochastic,
  title={Stochastic circuits for real-time image-processing applications},
  author={Alaghi, Armin and Li, Cheng and Hayes, John P},
  booktitle={Proceedings of the 50th Annual Design Automation Conference},
  pages={1--6},
  year={2013}
}

@inproceedings{alaghi2012spectral,
  title={A spectral transform approach to stochastic circuits},
  author={Alaghi, Armin and Hayes, John P},
  booktitle={2012 IEEE 30th International Conference on Computer Design (ICCD)},
  pages={315--321},
  year={2012},
  organization={IEEE}
}

@article{canny1986edge,
  title={A computational approach to edge detection},
  author={Canny, John},
  journal={IEEE Transactions on pattern analysis and machine intelligence},
  number={6},
  pages={679--698},
  year={2009},
  publisher={Ieee}
}

@inproceedings{li2011image,
  title={Using stochastic computing to implement digital image processing algorithms},
  author={Li, Peng and Lilja, David J},
  booktitle={2011 IEEE 29th international conference on computer design (ICCD)},
  pages={154--161},
  year={2011},
  organization={IEEE}
}

@inproceedings{suvizi2024exploring,
  title={Exploring in-memory accelerators and fpgas for latency-sensitive dnn inference on edge servers},
  author={Suvizi, Ali and Subramaniam, Suresh and Lan, Tian and Venkataramani, Guru},
  booktitle={2024 IEEE Cloud Summit},
  pages={1--6},
  year={2024},
  organization={IEEE}
}

@inproceedings{suvizi2025auto,
  title={Auto-Healer: Self-Healing Hardware for Perception Stage Faults in Autonomous Driving Systems},
  author={Suvizi, Ali and Venkataramani, Guru},
  booktitle={Proceedings of the 39th ACM International Conference on Supercomputing},
  pages={1064--1078},
  year={2025}
}

@article{adelkhani2025lpcr,
  title={LPCR: Information-Theory Based Low-Power Code Reordering for Serial Links in Network-on-Ch},
  author={Adelkhani, Morteza and Suvizi, Ali and Arzaghi, Farzaneh and Zamani, Sara and Salehi, Mohammad and Shafique, Muhammad},
  journal={IEEE Transactions on Computer-Aided Design of Integrated Circuits and Systems},
  year={2025},
  publisher={IEEE}
}

@article{suvizi2025parallel,
  title={A parallel computational approach for energy-efficient hydraulic analysis of water distribution networks using learning automata},
  author={Suvizi, Ali and Ahmadi, Ruhollah and Zamani, Morteza Saheb and Meybodi, Mohammad Reza},
  journal={Sustainable Computing: Informatics and Systems},
  volume={46},
  pages={101135},
  year={2025},
  publisher={Elsevier}
}

\end{document}